\input harvmac
%\draftmode
\let\includefigures=\iftrue
\let\useblackboard=\iftrue
\newfam\black 

%Figure Stuff
\includefigures
\message{If you do not have epsf.tex (to include figures),}
\message{change the option at the top of the tex file.}
\input epsf
\def\figin{\epsfcheck\figin}\def\figins{\epsfcheck\figins}
\def\epsfcheck{\ifx\epsfbox\UnDeFiNeD
\message{(NO epsf.tex, FIGURES WILL BE IGNORED)}
\gdef\figin##1{\vskip2in}\gdef\figins##1{\hskip.5in}% blank space instead
\else\message{(FIGURES WILL BE INCLUDED)}%
\gdef\figin##1{##1}\gdef\figinbs##1{##1}\fi}
\def\DefWarn#1{}
\def\figinsert{\goodbreak\midinsert}
\def\ifig#1#2#3{\DefWarn#1\xdef#1{fig.~\the\figno}
\writedef{#1\leftbracket fig.\noexpand~\the\figno}%
\figinsert\figin{\centerline{#3}}\medskip\centerline{\vbox{
\baselineskip12pt\advance\hsize by -1truein
\noindent\footnotefont{\bf Fig.~\the\figno:} #2}}
%\bigskip
\endinsert\global\advance\figno by1}
%%%
\else
\def\ifig#1#2#3{\xdef#1{fig.~\the\figno}
\writedef{#1\leftbracket fig.\noexpand~\the\figno}%
%\figinsert\figin{\centerline{#3}}\medskip
%\centerline{\vbox{\baselineskip12pt
%\advance\hsize by -1truein\noindent
%\footnotefont{\bf Fig.~\the\figno:} #2}}
%\bigskip\endinsert
\global\advance\figno by1} \fi

\def\id{{1 \kern-.28em {\rm l}}}

\def\K3{{\bf K3}}
\def\journal#1&#2(#3){\unskip, \sl #1\ \bf #2 \rm(19#3) }
\def\andjournal#1&#2(#3){\sl #1~\bf #2 \rm (19#3) }

\def\bar{\overline}
\def\hat{\widehat}
\def\ie{{\it i.e.}}
\def\eg{{\it e.g.}}

\def\tilde{\widetilde}

\def\frac#1#2{{#1\over#2}}

\def\inbar{\,\vrule height1.5ex width.4pt depth0pt}
\def\IC{\relax\hbox{$\inbar\kern-.3em{\rm C}$}}
\def\IR{\relax{\rm I\kern-.18em R}}
\def\IZ{\relax{\rm I\kern-.18em Z}}

%
%%%%%%%%%%%%%%%%%%%%%%%%%%%%%%%%%%%%
%

%
\catcode`\@=11
\def\slash#1{\mathord{\mathpalette\c@ncel{#1}}}
\overfullrule=0pt

\def\NN{{\cal N}}

\def\WW{{\cal W}}

\def\underrel#1\over#2{\mathrel{\mathop{\kern\z@#1}\limits_{#2}}}

\catcode`\@=12

%%%%%%%%%%%%%%%%%%%%%%%%%%%%%%%%%%%%%%%%%%%%%%%%%%%%%%%%%%%%%%

%

%%%%%%%%%%%%%%%%%%%%%%%%%%%%%%%%%%%%%%%%%%%%%%%%%%%%%%%%%%%%%%
% new defs:

\def\ie{{\it i.e.}}
\def\eg{{\it e.g.}}

%\SeibergPQ
\lref\SeibergPQ{
  N.~Seiberg,
  ``Electric - magnetic duality in supersymmetric nonAbelian gauge theories,''
Nucl.\ Phys.\ B {\bf 435}, 129 (1995).
[hep-th/9411149].
%%CITATION = hep-th/9411149%%
}

%\KutasovNP
\lref\KutasovNP{
  D.~Kutasov and A.~Schwimmer,
  ``On duality in supersymmetric Yang-Mills theory,''
Phys.\ Lett.\ B {\bf 354}, 315 (1995).
[hep-th/9505004].
%%CITATION = hep-th/9505004%%
}

%\KutasovSS
\lref\KutasovSS{
  D.~Kutasov, A.~Schwimmer and N.~Seiberg,
  ``Chiral rings, singularity theory and electric - magnetic duality,''
Nucl.\ Phys.\ B {\bf 459}, 455 (1996).
[hep-th/9510222].
%%CITATION = hep-th/9510222%%
}

%\KutasovVE
\lref\KutasovVE{
  D.~Kutasov,
  ``A Comment on duality in N=1 supersymmetric nonAbelian gauge theories,''
Phys.\ Lett.\ B {\bf 351}, 230 (1995).
[hep-th/9503086].
%%CITATION = hep-th/9503086%%
}

%\IntriligatorJJ
\lref\IntriligatorJJ{
  K.~A.~Intriligator and B.~Wecht,
  ``The Exact superconformal R symmetry maximizes a,''
Nucl.\ Phys.\ B {\bf 667}, 183 (2003).
[hep-th/0304128].
%%CITATION = hep-th/0304128%%
}

%\KutasovIY
\lref\KutasovIY{
  D.~Kutasov, A.~Parnachev and D.~A.~Sahakyan,
  ``Central charges and U(1)(R) symmetries in N=1 superYang-Mills,''
JHEP {\bf 0311}, 013 (2003).
[hep-th/0308071].
%%CITATION = hep-th/0308071%%
}

%\BrodieVX
\lref\BrodieVX{
  J.~H.~Brodie,
  ``Duality in supersymmetric SU(N(c)) gauge theory with two adjoint chiral superfields,''
Nucl.\ Phys.\ B {\bf 478}, 123 (1996).
[hep-th/9605232].
%%CITATION = hep-th/9605232%%
}

%\IntriligatorMI
\lref\IntriligatorMI{
  K.~A.~Intriligator and B.~Wecht,
  ``RG fixed points and flows in SQCD with adjoints,''
Nucl.\ Phys.\ B {\bf 677}, 223 (2004).
[hep-th/0309201].
%%CITATION = hep-th/0309201%%
}

%\GiveonSR
\lref\GiveonSR{
  A.~Giveon and D.~Kutasov,
  ``Brane dynamics and gauge theory,''
Rev.\ Mod.\ Phys.\  {\bf 71}, 983 (1999).
[hep-th/9802067].
%%CITATION = hep-th/9802067%%
}

%\GaddeLXA
\lref\GaddeLXA{
  A.~Gadde, S.~Gukov and P.~Putrov,
  ``(0,2) Trialities,''
[arXiv:1310.0818 [hep-th]].
%%CITATION = CALT-68-2862%%
}

%\BeniniCZ
\lref\BeniniCZ{
  F.~Benini and N.~Bobev,
  ``Exact two-dimensional superconformal R-symmetry and c-extremization,''
Phys.\ Rev.\ Lett.\  {\bf 110}, no. 6, 061601 (2013).
[arXiv:1211.4030 [hep-th]].
%%CITATION = arXiv:1211.4030%%
}
%\BeniniCDA
\lref\BeniniCDA{
  F.~Benini and N.~Bobev,
  ``Two-dimensional SCFTs from wrapped branes and c-extremization,''
JHEP {\bf 1306}, 005 (2013).
[arXiv:1302.4451 [hep-th]].
%%CITATION = arXiv:1302.4451%%
}

%\KutasovUX
\lref\KutasovUX{
  D.~Kutasov,
  ``New results on the 'a theorem' in four-dimensional supersymmetric field theory,''
[hep-th/0312098].
%%CITATION = hep-th/0312098%%
}

%\ErkalSH
\lref\ErkalSH{
  D.~Erkal and D.~Kutasov,
  ``a-Maximization, Global Symmetries and RG Flows,''
[arXiv:1007.2176 [hep-th]].
%%CITATION = arXiv:1007.2176%%
}

%\KlebanovYA
\lref\KlebanovYA{
  I.~R.~Klebanov and J.~M.~Maldacena,
  ``Superconformal gauge theories and non-critical superstrings,''
Int.\ J.\ Mod.\ Phys.\ A {\bf 19}, 5003 (2004).
[hep-th/0409133].
%%CITATION = hep-th/0409133%%
}

%\CurtoGE
\lref\CurtoGE{
  C.~Curto,
  ``Matrix model superpotentials and ADE singularities,''
Adv.\ Theor.\ Math.\ Phys.\  {\bf 12} (2008).
[hep-th/0612172].
%%CITATION = hep-th/0612172%%
}

%\IntriligatorAX
\lref\IntriligatorAX{
  K.~A.~Intriligator, R.~G.~Leigh and M.~J.~Strassler,
  ``New examples of duality in chiral and nonchiral supersymmetric gauge theories,''
Nucl.\ Phys.\ B {\bf 456}, 567 (1995).
[hep-th/9506148].
%%CITATION = hep-th/9506148%%
}

%%%%%%%%%%%%%%%%%%%%%%%%%%%%%%%%%%%%%%%%%%%%%%%%%
\Title{}
{\vbox{\centerline{Exceptional $N=1$ Duality}
\bigskip
%\centerline{..}
}}
\bigskip

\centerline{\it David Kutasov and Jennifer Lin}
\bigskip
%\smallskip
\centerline{EFI and Department of Physics, University of
Chicago} \centerline{5640 S. Ellis Av., Chicago, IL 60637, USA }
\smallskip

\vglue .3cm

\bigskip

\let\includefigures=\iftrue
\bigskip
\noindent 
Four dimensional $\NN=1$ supersymmetric gauge theory with gauge group $SU(N_c)$  and matter in the adjoint and fundamental representations gives rise to a series of fixed points with an ADE classification. The A and D series exhibit generalizations of Seiberg duality. We propose a similar duality for the $E_7$ theory.

\bigskip

\Date{}

\newsec{Introduction}

Shortly after Seiberg's work on the infrared behavior of $\NN=1$ supersymmetric QCD, and in particular his discovery of strong-weak coupling duality  in this theory \SeibergPQ, it was pointed out \refs{\KutasovVE\KutasovNP-\KutasovSS} that there is an infinite family of generalizations of  SQCD that has similar properties. These theories have gauge group $SU(N_c)$, $N_f$ flavors of chiral superfields $Q, \tilde Q$ that transform in the (anti) fundamental representation of the gauge group and a chiral superfield $X$ that transforms in the adjoint representation, with superpotential 
\eqn\aksup{\WW=s_0{\rm Tr} X^{k+1}\,.}
Here $k$ is a positive integer, and $s_0$ is a coupling. Naively, this coupling is irrelevant for $k>2$ and thus flows to zero in the IR. However it was argued in  \refs{\KutasovVE\KutasovNP-\KutasovSS} that for sufficiently small $N_f$ it actually influences the infrared behavior for all $k$, presumably because the quantum scaling dimension of the operator \aksup\ is reduced by the gauge interaction. The detailed mechanism for this was not understood until much later, but these theories were conjectured to have the following properties: 

\item{(1)} A stable supersymmetric vacuum for
\eqn\stelak{N_c\le kN_f\,.}
\item{(2)} A dual description in terms of a ``magnetic'' theory with gauge group $SU(kN_f-N_c)$, $N_f$ chiral superfields in the (anti) fundamental representation $q_i$, $\tilde q^i$, an adjoint field $\hat X$, and $k$ gauge singlets $M_j$, $j= 1, \cdots, k$, which transform in the bifundamental representation of the $SU(N_f)\times SU(N_f)$ flavor group.  The magnetic superpotential takes the form 
\eqn\mpot{
\WW \sim \Tr \hat X^{k+1}+\sum_{j=1}^k M_j\tilde q \hat X^{k-j}q  
}
where we omitted the coefficients of the different terms. The duality relates electric and magnetic chiral operators, 
\eqn\mmap{
 \tilde Q X^{j-1}Q \leftrightarrow M_j ,\;\;  {\rm Tr} X^j\leftrightarrow {\rm Tr}\hat X^j.
}
For $k=1$, the electric and magnetic adjoint fields $X$, $\hat X$ are massive, and the duality of \refs{\KutasovVE\KutasovNP-\KutasovSS} reduces to that of \SeibergPQ.
\item{(3)} The infrared behavior of these theories appears to be related to the study of mathematical singularities, a point of view that was particularly helpful when analyzing deformations of the superpotential \aksup\ \KutasovSS. 

\noindent
The last point was further developed in \BrodieVX. Viewing the superpotential \aksup\ as corresponding to an $A_k$ singularity, J. Brodie asked what happens if one replaces it with a $D_{k+2}$ one, 
\eqn\dksup{\WW\sim{\rm Tr} \left(X^{k+1}+XY^2\right)\,.}
He found a very similar structure to the $A_k$ case. There is again a lower bound on the number of flavors for which a stable supersymmetric vacuum exists, 
\eqn\steldk{N_c\leq 3kN_f}
and a dual description in terms of a magnetic theory with gauge group $SU(3kN_f-N_c)$ with the same charged matter, coupled to $3k$ singlet mesons 
\eqn\mlj{M_{lj}=\tilde Q X^{l-1}Y^{j-1}Q\;;\;\;\; l=1,\cdots, k; \;\;\; j=1,2,3}
via the superpotential
\eqn\spotl{
\WW \sim \Tr \hat X^{k+1} + \Tr \hat X\hat Y^2 + \sum_{\ell = 1}^k \sum_{j=1}^3 M_{\ell j}\tilde q \hat X^{k-\ell}\hat Y^{3-j} q\;.
}
This example includes two new elements compared to the $A_k$ case. One involves the matrix nature of the adjoint superfields. Although the superpotentials \aksup, \dksup\ look like the corresponding potential functions in singularity theory, they are functions of $N_c\times N_c$ matrices rather than single variables. In the $A_k$ case this distinction does not play a major role, since one can use the gauge symmetry and D-term constraints to diagonalize $X$, and view the superpotential \aksup\ as a function of its eigenvalues. The $D$-series involves two massless adjoints, $X$ and $Y$, and while one can use the above constraints to diagonalize one of them, one cannot diagonalize both at the same time. Thus, the $D$-series is the first case in which the matrix nature of the variables appearing in the superpotential plays a non-trivial role. 

The second new element in the work of \BrodieVX\ is the notion of quantum constraints on the chiral ring. Such constraints appeared already in the $A_k$ case (see \eg\ \KutasovSS), but they play a more central role in the $D$-series. Since similar constraints will feature prominently in our discussion below, we next briefly review the  main idea. 

The F-term constraints of the superpotential \dksup\ are\foot{Here and below we often neglect the contributions of Lagrange multipliers enforcing the tracelessness  of $X$, $Y$, which do not change the qualitative structure of what follows. We also pick a convenient relative normalization of the fields $X$ and $Y$.}
\eqn\ftermdk{X^k=Y^2\;; \qquad \{X,Y\}=0\,.}
Chiral operators are constructed from dressed quarks, $\Theta Q$, where $\Theta=\Theta(X,Y)$ is a polynomial in the adjoint fields, which satisfies the constraints \ftermdk. Superficially, these constraints lead to the infinite set
\eqn\formtheta{\Theta_{lj}=X^{l-1} Y^{j-1}\;;\;\;\; l=1,\cdots, k\;;\;\;\; j=1,2,\cdots.}
For odd $k$ the set \formtheta\ is actually further truncated to a finite one, since $Y^3=0$. Indeed, using the F-term constraints \ftermdk\ one has $Y^3=Y\cdot Y^2=Y\cdot X^k=-X^k\cdot Y=-Y^3$. Thus, the index $j$ in \formtheta\ runs only over the values $j=1,2,3$, in agreement with the fact that Brodie's duality only requires mesons with these quantum numbers \mlj, and baryons made of the corresponding truncated set of dressed quarks. 

For even $k$ this truncation appears to be absent, which is puzzling since the duality of \BrodieVX\ is expected to be valid for both even and odd $k$ (\eg\ because one can flow from odd to even $k$ by deforming the adjoint superpotential by relevant operators).  The solution to this conundrum proposed in \BrodieVX\ was that for even $k$ the constraint $Y^3=0$ appears quantum mechanically, so that the truncation to $j\le 3$ in \formtheta\ is the same for even and odd $k$ in the quantum theory, but not in the classical one. 

The origin of this quantum constraint in theories with even $k$ is not well understood. This is related to the fact that the vacuum structure of the theory with a general superpotential $W(X,Y)$ obtained by a relevant deformation of the $D_{k+2}$ superpotential \dksup\ is not fully understood either. For the $A_k$ case this analysis is easier, essentially because the single matrix $X$ can be diagonalized \refs{\KutasovVE\KutasovNP-\KutasovSS}, while for the $D$-series the non-abelian structure comes into play.

The understanding of RG flow in theories of the sort described above improved significantly with the advent of $a$-maximization \IntriligatorJJ. In particular, it was shown in \refs{\IntriligatorJJ,\KutasovIY} that the gauge theory with one adjoint superfield $X$ and no superpotential indeed has the property anticipated in \KutasovNP, that as  $N_c/N_f$ increases, the dimension of the chiral operator \aksup\ decreases in such a way that eventually it becomes relevant for all $k(<N_c)$. It was also shown in these papers that the properties of adjoint SQCD are consistent with the dualities of 
\refs{\KutasovVE\KutasovNP-\KutasovSS} and with the $a$-theorem. 

An important step in uncovering the ADE structure underlying the results of \refs{\KutasovVE\KutasovNP\KutasovSS-\BrodieVX} was taken in \IntriligatorMI. These authors used the techniques of \refs{\IntriligatorJJ,\KutasovIY} to classify all possible non-trivial fixed points of $\NN=1$ supersymmetric $SU(N_c)$ gauge theory with $N_f$ fundamentals $Q^i$, $\tilde Q_i$ and $N_a$ adjoints $X_\alpha$ that preserve the global $SU(N_f)\times SU(N_f)$ symmetry acting on the quarks. For $N_a>3$ the gauge theory is not asymptotically free and thus is expected to be trivial in the infrared. For $N_a=3$ interacting theories can only occur at $N_f=0$ (for the same reason), which from the general perspective is an isolated case. Thus, to have a non-trivial infrared behavior for non-zero $N_f$ one must take $N_a=2$ (or smaller). 

The authors of \IntriligatorMI\  considered models with two adjoint chiral superfields $X$ and $Y$,with superpotential $\WW=W(X,Y)$, and a tunable number of fundamentals $N_f$. Interestingly, they found that non-trivial fixed points correspond to superpotentials with an ADE structure, 
\eqn\rgfp{
\matrix{
 \hat O & W_{\hat O} = 0 \cr
 \hat A & \;\;\;\;\;\;\;W_{\hat A} = \Tr Y^2 \cr
 \hat D & \;\;\;\;\;\;\;\;\;\;W_{\hat D} = \Tr XY^2\cr
 \hat E & \;\;\;\;\;\;\;W_{\hat E} = \Tr Y^3 \cr
 A_k & \;\;\;\;\;\;\;\;\;\;\;\;\;\;\;\;\;\;\;\;\;\;\;\;W_{A_k} = \Tr(X^{k+1}+Y^2) \cr
 D_{k+2} & \;\;\;\;\;\;\;\;\;\;\;\;\;\;\;\;\;\;\;\;\;\;\;\;\;\;\;\;\;\;\;W_{D_{k+2}} = \Tr (X^{k+1}+XY^2) \cr
 E_6 & \;\;\;\;\;\;\;\;\;\;\;\;\;\;\;\;\;\;\;\;W_{E_6} = \Tr (Y^3 + X^4) \cr
 E_7 & \;\;\;\;\;\;\;\;\;\;\;\;\;\;\;\;\;\;\;\;\;\;\;W_{E_7} = \Tr (Y^3 + YX^3) \cr
 E_8 & \;\;\;\;\;\;\;\;\;\;\;\;\;\;\;\;\;\;\;\;\;\;W_{E_8} = \Tr(Y^3 + X^5)\,.
}}
These models naturally split into two classes. The first four $(\hat O, \hat A, \hat D, \hat E)$ are fixed points that exist for all $N_f$ satisfying the asymptotic freedom bound,\foot{The $\hat A$ theory can be thought of as having one adjoint superfield, $X$, and thus is asymptotically free for $N_f<2N_c$.} $N_f<N_c$, and can be thought of as UV ancestors of the rest. We will not discuss them further here. The last five have an ADE structure very reminiscent of that of mathematical singularities. 

The $A_k$ and $D_{k+2}$ theories in \rgfp\ were discussed above. The exceptional ones are new, and much about them remains mysterious. In particular:
\item{(1)} The $A$ and $D$ series fixed points only exist when the number of flavors is above a certain critical value, $N_f\ge N_f^{(\rm cr)}$, \stelak, \steldk. As we discuss below, there are reasons to believe that the same is true for the exceptional theories, but the bound is not known. 
\item{(2)} The $D_k$ models with $k$ even are believed to have the property that the classical ring \formtheta\ receives an important quantum correction ($Y^3=0)$. There are reasons to believe that the same is true for the exceptional theories, but the form of the corrections is not known. 
\item{(3)} In the $A$ and $D$ series models the features (1) and (2) are  closely related to duality. $N_f^{(\rm cr)}$ is the number of flavors for which the magnetic quarks disappear, and the quantum corrected set of dressed quarks $\Theta_jQ$ determines the spectrum of singlet mesons in the magnetic theory, as well as the spectrum of baryons. Thus, it is natural to ask whether the region $N_f\simeq N_f^{(\rm cr)}$ of the $E_k$ theories is better described in terms of a dual description. 

\noindent
In this paper we discuss the exceptional theories, \rgfp. We propose answers to questions (1) -- (3) above for the $E_7$ case, and discuss briefly the $E_6$ and $E_8$ theories. The hope is that a better understanding of the ADE theories will advance our understanding of Seiberg duality, and in particular of the quantum constraints on the chiral ring mentioned above.

The plan of the paper is the following. In section 2 we discuss some properties of the $E_7$ theory of \IntriligatorMI. We present the classical chiral ring of this theory and argue that its description in terms of the original, UV, variables must break down in the infrared for a sufficiently small number of flavors. We propose that this breakdown is related to a quantum constraint on the adjoint chiral superfields $X$, $Y$, which truncates the chiral spectrum in a way similar to that encountered in the $D_k$ theories with even $k$. 

In section 3 we show that this constraint is compatible with the existence of a weakly coupled dual description of the dynamics of the $E_7$ theory in the region of $(N_f,N_c)$ where the original (UV) description breaks down. The dual theory has gauge group $SU(30N_f-N_c)$, and matter content similar to that found in the dual descriptions of the $A_k$ and $D_k$ theories \refs{\KutasovVE\KutasovNP\KutasovSS-\BrodieVX}. This duality satisfies some rather non-trivial operator matching and `t Hooft anomaly matching constraints.  In section 4 we briefly comment on the remaining ($E_6$ and $E_8$) cases, leaving a more complete understanding of them to future work. 

\newsec{$E_7$ -- basic properties}

Looking back at \rgfp\ we see that the superpotential for the adjoint superfields is in this case 
\eqn\spoteseven{
\WW= \Tr\left( s_1Y^3 +s_2 Y X^3 \right)
}
where $s_1$, $s_2$ are couplings whose RG evolution depends on $N_f$, $N_c$. It is convenient to define the parameter \refs{\KutasovIY,\IntriligatorMI}
\eqn\defxxx{x={N_c\over N_f}}
which determines the strength of gauge interactions at long distances. It is of course a discrete parameter, that takes rational values; one can study the theory in the Veneziano limit $N_f,N_c\to\infty$, $x$ fixed, in which $x$ becomes continuous. This simplifies some of the formulae, and is not expected to make a qualitative difference in the dynamics. We will mostly work with general finite $N_f$, $N_c$; some of the numerical results below are stated in the Veneziano limit. 

Since we are interested in interacting IR fixed points, we will study the theory \spoteseven\ in the asymptotically free range $x>1$. As discussed in \IntriligatorMI, for all $x$ in this range, the coupling $s_1$ in the superpotential \spoteseven\ is relevant;  turning it on drives the theory to the $\hat E$ fixed point in \rgfp. The coupling $s_2$ can be relevant or not, depending on the R-charge of the operator ${\rm Tr} YX^3$ at the $\hat E$ fixed point. This problem can be addressed using $a$-maximization; one finds \IntriligatorMI\ that in the Veneziano limit this coupling is relevant for $x>x^{\rm min}\simeq 4.12$. Thus, for $1<x\le x^{\rm min}$, the $E_7$ fixed point coincides with the $\hat E$ one, while for larger $x$ the two are distinct. 

The $E_7$ fixed point, when it exists, has a global symmetry familiar from the $A$ and $D$ series models,  $SU(N_f) \times SU(N_f) \times U(1)_B \times U(1)_R$ under which the chiral superfields transform as follows:
\eqn\tfe{\eqalign{
Q \quad &\quad (N_f, 1, 1, 1 - \frac x 9) \cr
\tilde Q \quad &\quad (1, \bar{N_f}, -1, 1 - \frac x 9)\cr
X \quad &\quad (1, 1, 0, \frac 49) \cr
Y \quad &\quad (1, 1, 0, \frac 23).
}}
The superpotential \spoteseven\ leads to a truncation of the chiral ring. The equations of motion for $X$ and $Y$ set 
\eqn\cche{\eqalign{
& Y^2 = X^3  \cr
& X^2Y+XYX+YX^2 = 0
}}
where we neglected D-terms and chose a convenient relative normalization of $X$ and $Y$ (by choosing an appropriate normalization of the Kahler potential). As in the $A$ and $D$ series, we expect  an important role to be played by  the dressed quarks $\Theta(X,Y)Q$, where $\Theta(X,Y)$ is an arbitrary polynomial in $X$ and $Y$ modulo the relations \cche. These objects are the building blocks of  gauge invariant chiral operators (mesons and baryons). 

The dressed quarks can be constructed by systematically multiplying the $Q$'s (from the left) by $X$ and $Y$. This gives
\eqn\qquarks{\eqalign{
 \Theta_{(1,n)} &= X^n  \cr
\Theta_{(2,n)} &= YX^n  \cr
\Theta_{(3,n)} &= XYX^n  \cr
\Theta_{(4,n)} &= YXYX^n 
}}
where $n=0,1,2,\cdots$. Multiplying the operators \qquarks\ further by $X$ and $Y$ does not give anything new. Indeed, the chiral ring relations \cche\ lead to
\eqn\xqquarks{\eqalign{
X \Theta_{(1,n)} &= \Theta_{(1,n+1)} \cr
X \Theta_{(2,n)} &= \Theta_{(3,n)} \cr
X \Theta_{(3,n)} &= -\Theta_{(3,n+1)} -\Theta_{(2,n+2)}\cr
X \Theta_{(4,n)} &= \Theta_{(4, n+1)} \cr
Y \Theta_{(1,n)} &= \Theta_{(2,n)} \cr
Y \Theta_{(2,n)} &= \Theta_{(1,n+3)} \cr
Y \Theta_{(3,n)} &= \Theta_{(4,n)} \cr
Y \Theta_{(4,n)} &= \Theta_{(3,n+3)}.
}}
One can arrange the $\Theta$'s by R-charge, using the $U(1)_R$ charges in \tfe. The possible R-charges take the form $R={2\over9}N$, with $N=0,1,2,\cdots$. One can think of the $\Theta$'s as quasi-homogenous polynomials of degree $N$ in $X$, $Y$, with $X$ and $Y$ assigned degrees two and three, respectively. Beyond the first few levels, there are two operators at a given level, 
\eqn\Nevenodd{\eqalign{N=2k:&\qquad X^k,\,\, YXY X^{k-4}\cr
N=2k-1:&\qquad YX^{k-2},\,\,XYX^{k-3}\,.}}
For low $N$ there are fewer  operators: for $N=0,2,3,4,6$ there is one operator at each level. For $N=1$ there are none. 

The set of dressed quarks $\Theta_{(l,n)}$ \qquarks\ appears to give chiral meson operators of the form 
\eqn\defmln{M_{ln}=\tilde Q\Theta_{(l,n)} Q;\;\;\; l=1,2,3,4;\;\; n=0,1,2,\cdots}
These operators have R-charges 
\eqn\rmln{R_{ln}=2\left(1-{x\over9}\right)+{2\over9} N_{ln}}
with the $N_{ln}$ listed in \Nevenodd. As $x$ increases (\ie\ $N_f$ decreases), the R-charge \rmln\ decreases. At $x=N_{ln}$, the operator $M_{ln}$ becomes marginal, and shortly after that it goes below the unitarity bound $R=2/3$. At that point, something in our description must break down. As discussed in \KutasovIY, the nature of this breakdown varies from theory to theory. The mildest effect is the decoupling of the operator that violates the unitarity bound. This effect can be taken into account by allowing the superconformal R-symmetry of the infrared SCFT to mix with the emergent $U(1)$ flavor symmetry that only acts on the free field $M_{ln}$ \KutasovIY. 

In some cases, the modification of the theory at strong coupling is more severe, and involves a breakdown of the UV description of the whole theory, rather than merely the decoupling of particular operators. The picture proposed for this phenomenon in \refs{\KutasovIY,\KutasovUX,\ErkalSH} was the following. The description of the theory as a relevant deformation of a free UV theory provides a good description of a patch in the space of theories. As $x$ increases, we eventually leave this patch in the infrared, and need to look for a different description. That description is often provided by strong/weak coupling duality. 

In the ADE theories \rgfp, it is believed that the first four do not suffer from the more severe version of the breakdown, and can be studied using the UV description for all $x$ (\ie\ all $N_f$). The $A_k$ and $D_k$ theories are different. Extrapolating the UV description, even taking into account the decoupling of meson fields that reach R-charge $2/3$, fails for sufficiently large $x$; in that regime the correct description is given by the dual theory. 

It is natural to expect the same to happen for the exceptional theories. To test this for the $E_7$ theory, we calculate the central charge $a(x)$ assuming that the UV description (corrected by the decoupling of free mesons) is valid, following the techniques of \refs{\KutasovIY,\IntriligatorMI}.  We find the curve given in figure 1. According to the $a$-theorem, this curve must be monotonically decreasing, since increasing $x$ (with fixed $N_c$) means decreasing $N_f$, which can be achieved by adding relevant deformations (mass terms) to the Lagrangian. However, we see that the curve in fact has a minimum around $x\sim 27$, after which it violates the $a$-theorem. This suggests that the assumption that the UV variables provide an accurate description all the way to the infrared fails at large $x$. 

\bigskip

\ifig\loc{The plot of the function $x^{-2}a(x)/N_f^2$ in the UV description of the $E_7$ theory, taking into account the decoupling of free mesons.}
{\epsfxsize4in\epsfbox{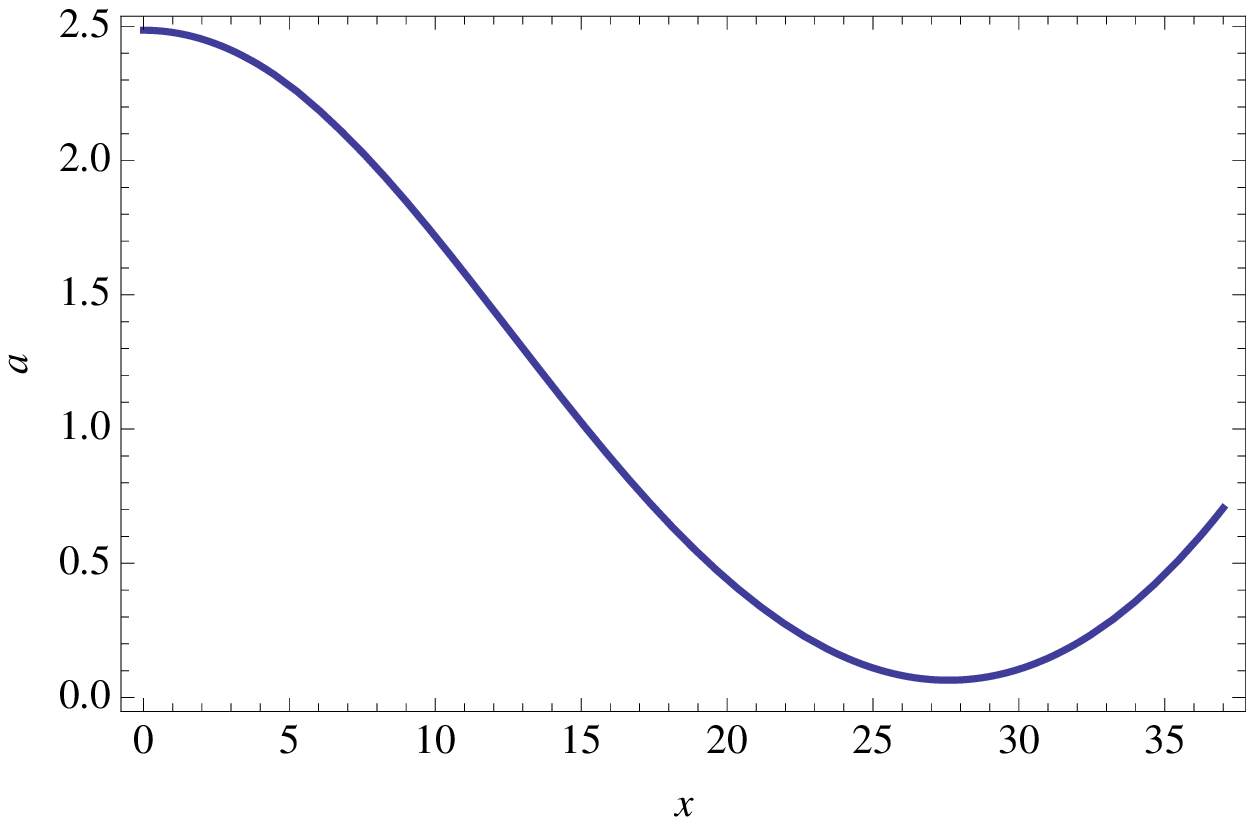}}

\bigskip

We note in passing that the curve of figure 1 receives a further important modification from baryons reaching the unitarity bound. If we keep all dressed quarks $\Theta_iQ$ \qquarks, the R-charge of the lowest baryon as a function of $x$ is
\eqn\bbarx{
N_f^{-1}R_B(x) = x R(Q) + \sum_{i=1}^x R(\Theta_i)
}
where for simplicity we took $x$ to be integer,  and the sum runs over the operators \Nevenodd\ of lowest R-charge. $R_B(x)$ crosses the unitarity bound
around $x=29$, and after that point one needs to recalculate the curve of figure 1 taking into account the decoupling of the baryons. Since, as mentioned above, the description leading to this curve must break down before $x=27$, this is not particularly important, but it provides further evidence for the fact that something non-trivial must happen in this theory around $x\sim 25-30$.

We will assume that the modification in question is similar to that in the $A$ and $D$ series, \ie\ there exists a dual theory that provides a better description of the region where the $a$-theorem is superficially violated. Furthermore, we will assume that the qualitative structure of this dual theory is similar to that of the other cases, \ie\ the electric mesons \defmln\ appear in it as singlets of the magnetic gauge group, and are coupled to the charged magnetic fields via a superpotential of the general form \spotl. 

In order for this picture to make sense, it must be that the number of chiral electric meson operators is finite; we will denote it by $\alpha$. Thus, there must be a constraint on the dressed quarks, or equivalently on $\Theta_{(l,n)}$ \qquarks, which cuts down their number to a set of $\alpha$ independent operators. Such a constraint would have to be due to quantum effects. Fortunately, we have a precedent for this phenomenon -- the $D_k$ theories with even $k$ -- where exactly this is believed to happen, so we will proceed under this assumption here as well. 

The simplest scenario is that there is a single quantum constraint of the form $F(X,Y)Q=0$ where $F$ is a quasi-homogenous polynomial of   degree $n$, with $X$ assigned degree two and $Y$ degree three. We will see that this assumption gives a sensible picture, and the degree $n$ is uniquely determined. Thus, we will not consider more complicated scenarios.

The level at which constraints on the chiral operators $\Theta(X,Y)$ first appear is $n$. We can get constraints at higher level by multiplying the constraint $F(X,Y)=0$ from the left by $X$ and $Y$. This leads to the following pattern: at level $n+1$ there are no constraints, at levels $n+2$, $n+3$, $n+4$ one constraint, obtained by multiplying $F(X,Y)=0$ by $X$, $Y$ and $X^2$, respectively, at level $n+5$ two constraints, from $XY$ and $YX$, at level $n+6$ one constraint, from $Y^2=X^3$, and at level $n+7$ and on two constraints. Thus, the number of independent $\Theta(X,Y)$ at these levels follows the pattern: 
\eqn\pattern{\eqalign{
{\rm level:}&  \;\;\;(n,n+1,n+2,\cdots)\cr
{\rm number\; of\; states:}& \;\;\;(1,2,1,1,1,0,1,0,0,\cdots)
}}
Since there are no non-trivial $\Theta_i$ at a level larger than $n+6$, the total number of meson operators, $\alpha$, is rendered finite by the constraint. If the level $n$ is not too small $(n\ge 7)$ one finds 
\eqn\formalpha{\alpha=2n\,.}
We will next see that $n$ is uniquely determined by duality.

\newsec{$E_7$ -- duality}

To proceed, we will assume that the $E_7$ theory satisfies a generalization of the duality of the $A$ and $D$ series discussed above. We will take the gauge group to be $SU(\hat N_c)$, and the matter to consist of $N_f$ (anti) fundamentals $q_i$, $\tilde q^i$, adjoint fields $\hat X$, $\hat Y$, and a superpotential of the form 
\eqn\magsup{\WW\sim \Tr\left(\hat Y^3 +\hat Y \hat X^3 \right)+\sum_{j=1}^\alpha M_j\tilde q\hat\Theta_j q\,.
}
The transformation of the magnetic fields under the global symmetry (the analog of \tfe) is 
\eqn\mfields{\eqalign{
q \qquad & \qquad (\bar N_f, 1, {N_c\over \hat N_c}, 1-{\hat x\over 9}) \cr
\tilde q \qquad & \qquad (1, N_f, -{N_c\over \hat N_c}, 1-{\hat x\over 9}) \cr
\hat X \qquad & \qquad (1,1,0,\frac 49) \cr
\hat Y \qquad & \qquad (1,1,0,\frac 23) \cr
M_j \qquad & \qquad (N_f, \bar N_f, 0, R_j)\,. \cr
}}
As we explain below, on general grounds one must have 
\eqn\nchat{\hat N_c=\alpha N_f-N_c}
so that the magnetic coupling $\hat x$ takes the form (compare to \defxxx) $\hat x=\alpha-x$. The singlet mesons $M_j$, $j=1,\cdots,\alpha$, correspond to the electric gauge invariant fields  
\eqn\formmj{M_j\leftrightarrow \tilde Q\Theta_j Q}
where $\Theta_j$ are the polynomials in $X$, $Y$ discussed in the previous section. Thus, their R-charges are (see \rmln)
\eqn\rrmm{R_j=2\left(1-{x\over 9}\right)+{2\over9}N_j\,.}
Here $N_j$ is the degree of the quasi-homogenous polynomial $\Theta_j$, with $X$ assigned degree two and $Y$ degree three.

The meson terms in the magnetic superpotential \magsup\ pair each electric operator $\Theta_j$ made out of the adjoints $X$, $Y$ with a magnetic one $\hat\Theta_j$ made out of $\hat X$, $\hat Y$. The form of the superpotential implies that the two paired operators  satisfy 
\eqn\rrjj{N_j+\hat N_j=\alpha-9\,.} 
This means that operators from the bottom of the list \Nevenodd\ in the electric theory are paired with operators from the top of the list \pattern\ in the magnetic theory, and vice versa. Fortunately, the spectrum has the right structure for such a pairing. The unique operator at level $n+6$ is paired with the identity operator, the level $n+4$ operator is paired with $X$, etc. The R-charge constraint above then implies that 
\eqn\rchconst{n+6=\alpha-9}
which together with \formalpha\ gives $n=15$ and $\alpha=30$. Thus, duality relates the electric gauge group $SU(N_c)$ to the magnetic one $SU(30N_f-N_c)$.

We conclude that if a duality of the sort found in the $A$ and $D$ series is to exist in the $E_7$ theory, one must impose on the spectrum \Nevenodd\ a quantum constraint of the form 
\eqn\quantconst{aYX^6+bXYX^5=0}
where $a$, $b$ are constants that are not determined by the above considerations. This constraint truncates the infinite set of mesons to a finite set, which is (uniquely) consistent with such a duality.

An important check of duality in other cases is the matching of `t Hooft anomalies for the global currents. In the electric theory, the non-vanishing anomalies take the form 
\eqn\thae{\eqalign{
SU(N_f)^3 \quad & \quad  N_c d^{(3)}(N_f) \cr
SU(N_f)^2 U(1)_R \quad & \quad  -\frac{x}{9}N_c d^{(2)}(N_f) \cr 
SU(N_f)^2U(1)_B \quad & \quad N_c d^{(2)}(N_f) \cr
U(1)_R \quad & \quad - \frac 1 9 (N_c^2+1) \cr
U(1)_R^3 \quad & \quad \frac{577}{729}(N_c^2-1) -\frac{2}{729}\frac{N_c^4}{N_f^2} \cr
U(1)_B^2U(1)_R \quad & -\frac 2 9 N_c^2 \quad 
}}
where $d^{(3)}(N_f) \sim \Tr T^a\{T^b, T^c\}, d^{(2)}(N_f) \sim \Tr T^a T^b$, with the traces taken in the fundamental representation. 

The anomalies in the magnetic theory can be expressed in terms of $r_j$, the R-charges of  the operators $\Theta_j$ in \formmj.  Denoting $r_j=2N_j/9$, the spectrum we found above contains one operator each at $N=0,2,3,4,6,15,17,18,19,21$, and two operators for each of $N=5,7-14,16$. The operators with $N$ and $21-N$ are paired by the magnetic superpotential, as explained around \rrjj. 

The $SU(N_f)^3$ anomaly in the magnetic theory is $-\hat N_c+\alpha N_f$. Its matching with the first line of \thae\  is the origin of the condition \nchat. The matching of the other anomalies can be shown to reduce to the three conditions
\eqn\cfoure{\eqalign{
\sum r_j &= \frac{\alpha^2}{9} - \alpha \cr
\sum r_j^2 &= \frac{4\alpha^3}{243} - \frac{2\alpha^2}{9} - \frac{334\alpha}{243} \cr
\sum r_j^3 &= \frac{2\alpha^4}{729} - \frac{4\alpha^3}{81} - \frac{334\alpha^2}{729} + \frac{496\alpha}{81}~.
}}
In particular, the $SU(N_f)^2 U(1)_R$ and $U(1)_R$ anomalies follow from the first line of \cfoure, while the $U(1)_R^3$ one requires all three conditions. 
All anomalies in \thae\ that involve $U(1)_B$ are automatically satisfied for all $\alpha$. 

The first of the three equations in \cfoure\  actually follows from the form of the magnetic superpotential \magsup, which as we discussed leads to a pairing of different $\Theta_j$ into pairs satisfying the sum rule  \rrjj. Summing this sum rule over all pairs leads to the constraint on the first line of \cfoure. Thus, this constraint is automatically satisfied by the spectrum of $\Theta_j$ described above. The remaining two equations in \cfoure\ need to be checked and provide a (very) non-trivial check on our duality. One can check that they are in fact satisfied for the spectrum found above.

To complete the specification of the duality, we need to determine the mapping of the baryon operators between the electric and magnetic theories. This is done as in the $A$ and $D$ cases \refs{\KutasovVE\KutasovNP\KutasovSS-\BrodieVX}. The electric baryons can be constructed out of the dressed quarks, $Q_{(j)}=\Theta_j Q$, $j=1,\cdots,\alpha$:
\eqn\elecbar{B^{(l_1,\cdots,l_\alpha)}=Q_{(1)}^{l_1} \cdots Q_{(\alpha)}^{l_\alpha};\;\;\; \sum_{j=1}^\alpha l_j=N_c~.}
The total number of baryon operators is
\eqn\totbarele{\sum_{\{l_j\}}{N_f\choose l_1}\cdots {N_f\choose l_\alpha}={\alpha N_f\choose N_c}\,.}
The map of baryons between the electric and magnetic theories is 
\eqn\elmagbar{B_{\rm el}^{(l_1,\cdots,l_\alpha)}\leftrightarrow B_{\rm mag}^{(\hat l_1,\cdots,\hat l_\alpha)};\;\;\; \hat l_j=N_f-l_j\,.
}
As in the $A$ and $D$ series dualities, it is non-trivial that the charge assignment \mfields\ that is necessary for anomaly matching is consistent with the map \elmagbar. 

Taking into account the duality, we are led to the following phase structure of the theory as a function of the coupling $x=N_c/N_f$:
\eqn\phs{\matrix{
x \leq 1  \qquad  \qquad  &{\rm free\,\,electric} \cr
1 < x < x^{\min}_{E_7}   \qquad  \qquad  &\hat E {\rm\,\,electric} \cr
x^{\min}_{E_7} < x < 30 - \hat x^{\min}_{E_7}  \qquad  \qquad  &{\rm conformal\,\,window} \cr
30-\hat x^{\min}_{E_7} < x < 29  \qquad  \qquad  &\hat E {\rm\,\,magnetic} \cr
29\leq x\le 30  \qquad  \qquad   &{\rm free\,\,magnetic}\cr
30<x \qquad\qquad &{\rm no\; vacuum}
}}
where $\hat x = 30-x$, and $x^{\min}_{E_7} \sim 4.12$ as discussed above. $\hat x^{\min}_{E_7}$ differs slightly from $x^{\min}_{E_7}$ because of the mesons and superpotential in the magnetic theory; solving for it we find that it is around $\hat x^{\min}_{E_7} \sim 3.89$. The duality also helps to resolve the problem with the $a$-theorem we found before (see figure 1). Taking it into account leads to the $a$-function plotted in figure 2. 

\bigskip

\ifig\loc{Our results modify the $a$ function of figure 1 to this curve. The dashed line describes the $\hat E$ and free magnetic phases. For $x>30$ the theory has no vacuum.}
{\epsfxsize4in\epsfbox{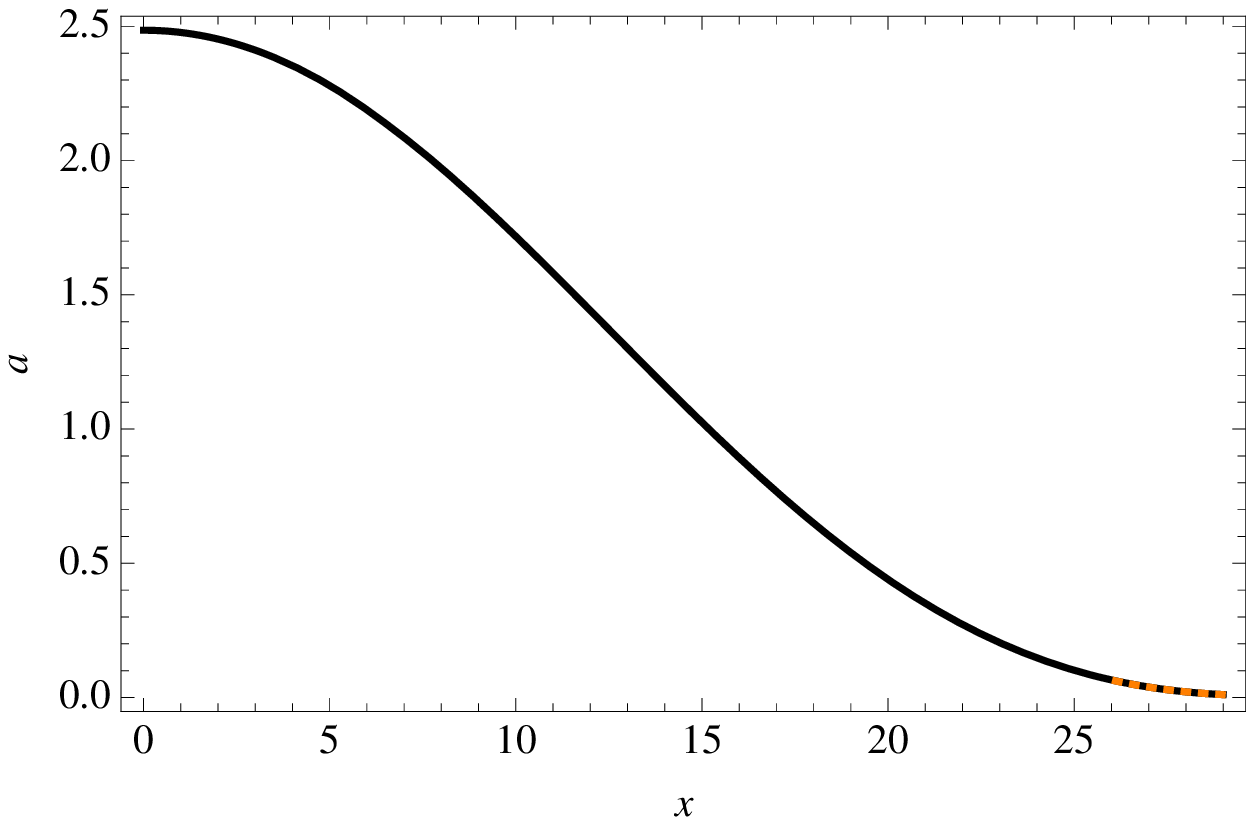}}

\bigskip

As in other cases, one can further test the duality we found by studying deformations. We will for the most part leave this to future work, but will finish this section with a few comments. One deformation is giving a mass to a flavor. In the electric theory this takes $(N_c,N_f)\to (N_c,N_f-1)$. In the magnetic theory this means $(\hat N_c,N_f)\to (\hat N_c-30,N_f-1)$. The mass term corresponds in the magnetic theory to adding to the superpotential the term $\delta W=m (M_1)_{N_f}^{N_f}$. Adding this to \magsup\ and varying w.r.t. the $M_j$ gives a non-zero expectation value for the magnetic meson fields, 
\eqn\vevmag{\langle\tilde q^{N_f}\hat\Theta_j q_{N_f}\rangle\sim m\delta_{j,1};\;\;\; j=1,\cdots, 30.}
Since the thirty operators $\hat\Theta_j$ are linearly independent, it is natural that the magnetic gauge group is broken to $SU(\hat N_c-30)$ by \vevmag. 

Other deformations correspond to going along the electric moduli space, and deforming the superpotential for $X$ and $Y$. The former should be straightforward to match between the electric and magnetic theories. The latter is not well understood already in the $D$-series, and would be very interesting to understand here, among other reasons since one can flow from the $E_7$ theory to the $E_6$ one which, as we will discuss next, is not well understood.

\newsec{Discussion}

There are many questions raised by our results. In the $E_7$ and $D_k$ theories with even $k$, it would be interesting to understand in a deeper way the origin of the quantum constraints on the chiral operators constructed out of the adjoint fields, $\Theta_j$. It is also important to obtain a good understanding of the vacuum structure in the space of deformations of these theories obtained by varying the adjoint superpotential $\WW(X,Y)$. 

Another set of questions involves the $E_6$ and $E_8$ theories. These theories seem to be somewhat different from the cases studied in previous work and in this paper. Consider, for example, the $E_6$ theory, whose superpotential is 
\eqn\wesix{\WW\sim{\rm Tr} \left(X^4+Y^3\right).} 
This superpotential is invariant under separate $SU(N_c)$ rotations of the two massless adjoints $X$ and $Y$, in contrast to that of the $E_7$ theory \spoteseven, and of the $D_k$ ones. A related feature is that the F-term equations 
\eqn\cch{
X^3 = Y^2 =0
}
do not couple $X$ and $Y$. This leads to a rich classical spectrum of dressed quarks $\Theta Q$, with 
\eqn\xy{
\Theta=X^{n_0} Y X^{n_1} Y \cdots Y X^{n_k}
}
where $n_0,n_k=0,1,2$ and the rest of the $n_j=1,2$.  

The transformation properties of the matter fields under the $SU(N_f) \times SU(N_f) \times U(1)_B \times U(1)_R$ global symmetry are 
\eqn\tf{\eqalign{
Q\quad &\quad (N_f, 1, 1, 1 - \frac x 6) \cr
\tilde Q \quad & \quad (1, \bar{N_f}, -1, 1 - \frac x 6) \cr
X \quad &\quad  (1, 1, 0, \frac 12) \cr
Y\quad &\quad (1,1,0, \frac 23)
}}
where $x = N_c / N_f$.
Thus, the R-charges of the operators \xy\ are
\eqn\dxy{
R(\Theta) = \sum_{j=0}^k \frac 12 n_j + \frac 23 k\,.
}
Clearly, the number of operators at a given value of the R-charge grows with the R-charge, unlike the $E_7$ case, where this number is equal to two \Nevenodd. This is due to the fact that the $E_6$ (and $E_8$)  F-term constraints do not relate different orderings of $X$ and $Y$, unlike the $E_7$ and $D_k$ ones. 

\bigskip

\ifig\loc{The plot of the function $x^{-2}a(x)/N_f^2$ in the UV description of the $E_6$ theory, taking into account the decoupling of free mesons.}
{\epsfxsize4in\epsfbox{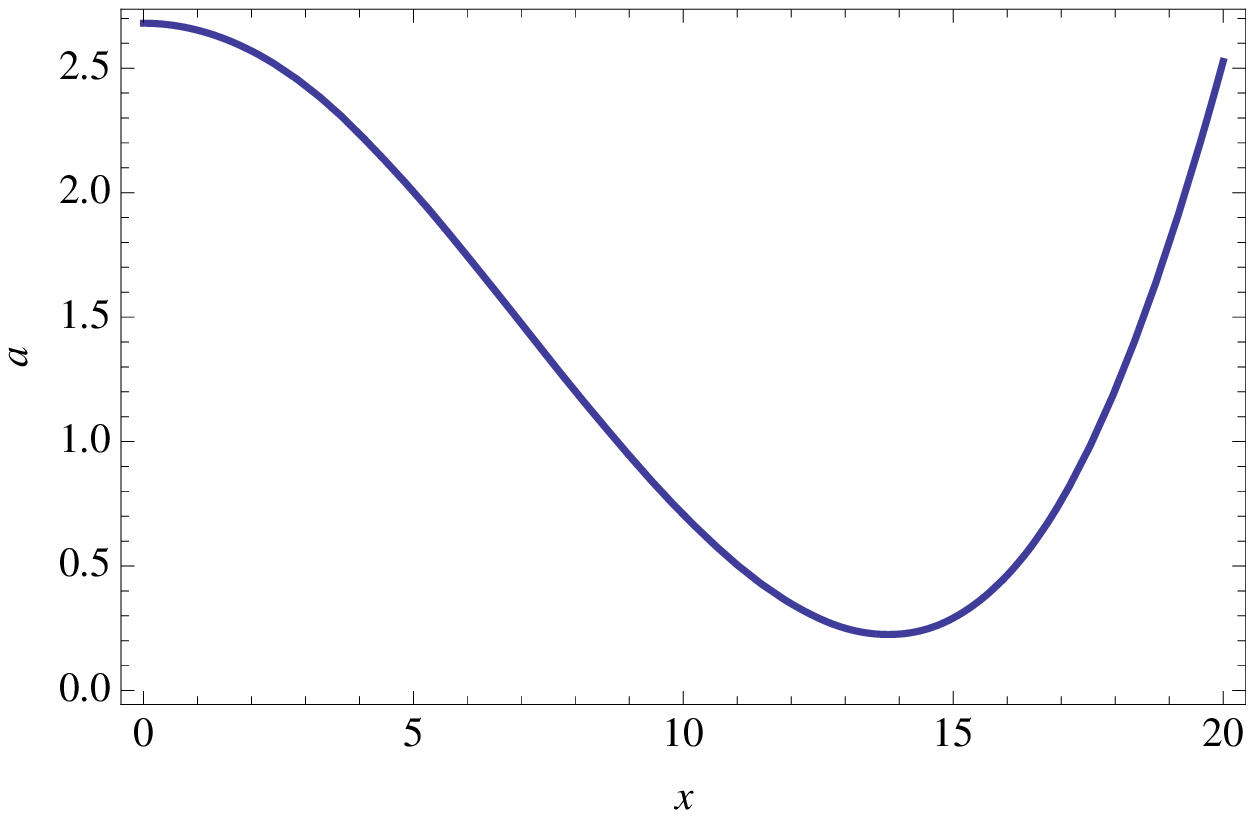}}

\bigskip

One can try to repeat the discussion of the $E_7$ theory for this case. The superpotential \wesix\ can be thought of as a perturbation of the $\hat E$ theory by the operator ${\rm Tr} X^4$. As shown in \IntriligatorMI, this perturbation becomes relevant at $x\simeq 2.55$. As $X$ increases, one again starts finding meson operators $\tilde Q\Theta Q$ whose R-charge naively drops below the unitarity bound. Assuming that the only effect of strong coupling is the decoupling of such mesons, as in the discussion leading to figure 1, leads to the $a$ function depicted in figure 3 (which reproduces and extends figure 21 in \IntriligatorMI). 

As there, the non-monotonicity of $a$ at $x\simeq 13.8$ suggests that the description of the $E_6$ theory in terms of a free UV fixed point breaks down in the IR around there. Given our discussion above, it is natural to expect a dual description with gauge group $SU(\alpha N_f-N_c)$, with $\alpha$ in the teens. Under some rather mild assumptions one can show that no such dual exists. Nevertheless, the theory must find a way to preserve unitarity; it would be interesting to see how it does that, perhaps by finding a weakly coupled description of the physics in the problematic region $x>13$.

The construction of \refs{\KutasovVE\KutasovNP-\KutasovSS} was generalized to a much larger class of theories with different gauge groups and matter representations in \IntriligatorAX. It would be interesting to generalize the discussion of \refs{\KutasovIY,\IntriligatorMI} and this paper to this class of theories. It would also be interesting to realize the ADE theories \rgfp\ as low energy theories on branes \GiveonSR, and describe them via gauge/gravity duality, \eg\ by generalizing the construction of \KlebanovYA. This may provide a deeper understanding of the origin of the ADE structure underlying these theories, the origin of the quantum constraints in the $E_7$ and $D_k$ theories with even $k$, and may shed light on the remaining cases ($E_6$ and $E_8$). There is also a potential relation to the discussion of \CurtoGE; it would be nice to understand it better.
 
\bigskip 
 
\noindent{\bf Acknowledgements}: We thank O. Aharony and K. Intriligator for discussions. This work was supported in part by DOE grant DE-FG02-13ER41958, NSF Grant No. PHYS-1066293 and the hospitality of the Aspen Center for Physics, and by the BSF -- American-Israel Bi-National Science Foundation. The work of JL was supported in part by an NSF Graduate Research Fellowship. DK thanks IPMU Tokyo, Tel Aviv University and the Hebrew University for hospitality at various stages of this work.

\listrefs

\bye